\documentclass[sigconf,nonacm]{acmart}

\usepackage{graphicx}
\usepackage{url}

\newcommand{\q}[1]{\lq\lq{}{}#1\rq\rq{}{}}
\newcommand{\sq}[1]{\lq{}#1\rq{}}

\begin{document}

\title[FastCat Catalogues]{FastCat Catalogues: Interactive Entity-based Exploratory Analysis of Archival Documents}

\author{Georgios Rinakakis}
\email{rinakakis1999@gmail.com}
\affiliation{%
  \institution{Information Systems Laboratory, ICS-FORTH \&\\Department of Computer Science, University of Crete}
  \city{Heraklion}
  \country{Greece}
}

\author{Kostas Petrakis}
\email{cpetrakis@ics.forth.gr}
\orcid{0000-0003-2788-526X}
\affiliation{%
  \institution{Information Systems Laboratory, ICS-FORTH}
  \city{Heraklion}
  \country{Greece}
}

\author{Yannis Tzitzikas}
\email{tzitzik@ics.forth.gr}
\orcid{0000-0001-8847-2130}
\affiliation{%
  \institution{Information Systems Laboratory, ICS-FORTH \&\\Department of Computer Science, University of Crete}
  \city{Heraklion}
  \country{Greece}
}

\author{Pavlos Fafalios}
\email{fafalios@ics.forth.gr}
\orcid{0000-0003-2788-526X}
\affiliation{%
  \institution{Information Systems Laboratory, ICS-FORTH}
  \city{Heraklion}
  \country{Greece}
}

\begin{abstract}
We describe FastCat Catalogues, a web application that supports researchers studying archival material, such as historians, in exploring and quantitatively analysing the data (transcripts) of archival documents. 
The application was designed based on real information needs provided by a large group of researchers,
makes use of JSON technology, and is configurable for use over any type of archival documents whose contents have been transcribed and exported in JSON format. 
The supported functionalities include 
a)~source- or record-specific entity browsing, 
b)~source-independent entity browsing, 
c)~data filtering, 
d)~inspection of provenance information, 
e)~data aggregation and visualisation in charts,
f)~table and chart data export for further (external) analysis.
The application is provided as open source and is currently used by historians in maritime history research.
\end{abstract}

%



\maketitle

\section{Introduction}
\label{sec:intro}
Archival research is a type of research which involves investigating and extracting evidence from original archival material, such as historical documents \cite{ventresca2017archival}.
These documents often have a repetitive structure, providing information about one or more \textit{categories of entities}, such as persons or locations. Examples include logbooks, payrolls, censuses, civil registers, employment records, etc. 

Researchers that study such documents usually start by first transcribing the important archival information in digital form. This enables them in performing exploratory analysis on the transcribed data, such as quantitative analysis of collective phenomena for drawing conclusions on possible impact factors~\cite{petrakis2020digitizing}. 

For example, consider the real use case of the SeaLiT project (ERC project in the field of maritime history)\footnote{\url{https://sealitproject.eu}}, which studies the transition from sail to steam navigation and its effects on seafaring populations in the Mediterranean and the Black Sea between the 1850s and the 1920s \cite{delis2020seafaring}.
Maritime historians have collected and transcribed a large and diverse set of archival documents ($>$600 records, 20 different types of sources, 5 languages), such as crew lists, payrolls, logbooks, and sailor registers, gathered from multiple authorities in different countries. 
These documents provide historical information for thousands of interconnected entities, including ships, captains, sailors, ship owners, departure ports, embarkation ports, residence locations, etc.

To effectively explore and analyse this amount of data, and extract useful information for their research, historians need interactive and intuitive-to-use interfaces that can support them in finding the desired information. 
To this end, in this paper we present FastCat Catalogues, a web-based system that allows the entity-based exploration and analysis of archival documents. 
The system has been designed by considering \textit{real} information needs provided by a large group of historians. The aim is to provide a system that can support historians in finding answers to their information needs.

For a selected entity (e.g. a ship), the user can inspect its connection to other entities (e.g. crew members, voyages, departure ports, arrival ports), or directly visit the original transcripts that mention this entity for validation or retrieval of further (contextual) information. The user can also group a list of displayed entities by one of their characteristics and visualise the result in a chart (such as grouping the ship's crew members by their residence location). 

Through a dedicated configuration model that allows defining the entities of interest and their relation to other entities per source type, the application can be configured for use over any type of archival documents whose transcripts are exported in JSON format. 

The application is available as open source\footnote{\label{gitlink}\url{https://github.com/isl/FastCat-Catalogues}
}
 and has been deployed considering data of the SeaLiT project. 
The deployment is publicly available in the following link:

\begin{center}
\url{https://catalogues.sealitproject.eu/}
\end{center}

The rest of this paper is organised as follows:
Section~\ref{sec:rw} discusses related work.
Section~\ref{sec:reqAn} describes the design requirements.
Section~\ref{sec:app} details the system's functionality, user interface (UI) and technology.
Section~\ref{sec:conf} presents the system's configuration model. 
Section~\ref{sec:usecase} discusses its evaluation and use in a real context.
Finally, Section~\ref{sec:conclusion} concludes the paper and outlines future work.

\section{Related Work and Innovation} 
\label{sec:rw}
The majority of existing works supports searching archival documents based on metadata and/or textual search on their contents. Such systems are usually based on search engines like Elasticsearch\footnote{\url{https://elastic.co}} or Apache Solr\footnote{\url{https://solr.apache.org}}, e.g.~\cite{pham2020building} and \cite{marciano2018archival}. 

The innovation of FastCat Catalogues compared to these systems lies on the fact that it allows browsing and exploring information about a set of \textit{highly-interrelated} entities (e.g. sailors, ships, ports, locations, etc., in the case of SeaLiT), where the type of the relation between the entities is highly dependant on the entities context in the archival documents (e.g. a location can be a departure port for a ship, a destination port, a port of call, etc.).

Another approach is to make use of semantic technologies for representing archival/historical data as a rich knowledge graph of linked data~\cite{hawkins2022archives,oldman2016zen,merono2015semantic,fafalios2022building}.
Then, user-friendly interfaces aim at offering an intuitive way to the end users for accessing and exploring the data (e.g.~\cite{kritsotakis2018assistive,oldman2018reshaping}). 
Such an approach allows specifying the semantic relations among the involved entities as well as linking the data to external datasets (if needed). 
Nevertheless, it requires extensive conceptual modeling work for representing the data ontologically, while offering a reliable, robust and efficient service over such semantic repositories is a challenging task. 
Semantifying the application is part of our future work, e.g. either by considering RDF triples as the input data, and/or by supporting the extraction of the displayed data to RDF.

\section{Requirement Analysis}
\label{sec:reqAn}

We have collected a set of more than 100 information needs (\textit{competency questions}) from a group of around twenty historians belonging to different research groups in five countries and working with archival documents in the context of the SeaLiT project. These information needs are directly related to the archival material, i.e. its analysis can provide answers to the information needs or important related evidence.
Indicative examples are:\footnote{The full list of gathered information needs is available at \url{https://users.ics.forth.gr/~fafalios/SeaLiT_Competency_Questions_InfoNeeds.pdf}} 
\begin{itemize}
    \item What are the places of construction of ships during a specific period?
    \item What are the most popular European destinations in different time periods of ships departed from ports in the Black Sea?
    \item How many ship owners per ship during a specific time period?
\end{itemize}

We tried to group the information needs in categories and used them as requirements for the design of the application. We identified the below four main categories: 
(i)~finding information about a particular entity, such as the birth date and place of a person; 
(ii)~retrieving a list of entities based on one or more properties of these entities, together with additional information about them, e.g. all persons having a specific residence location together with their birth date; 
(iii)~grouping a list of retrieved
entities based on some property or characteristic, e.g. grouping all retrieved persons by their profession; 
(iv)~finding comparative information related to some entities, e.g. number of persons employed by the organisation in different time periods, or voyage duration per type of ship.

The aim is to design a system that can support humanities researchers studying archival material in finding answers to these categories of information needs.

Another important requirement highlighted by the historians is the ability to find the provenance of any piece of displayed information, by enabling end users to directly visit the original transcript that contains the information. This is very important for historians because it allows finding additional contextual data quickly, but also for verification purposes.

\section{UI, Functionality and Technology}
\label{sec:app}

\begin{figure}[h]
    \centering
    \includegraphics[width=8.7cm]{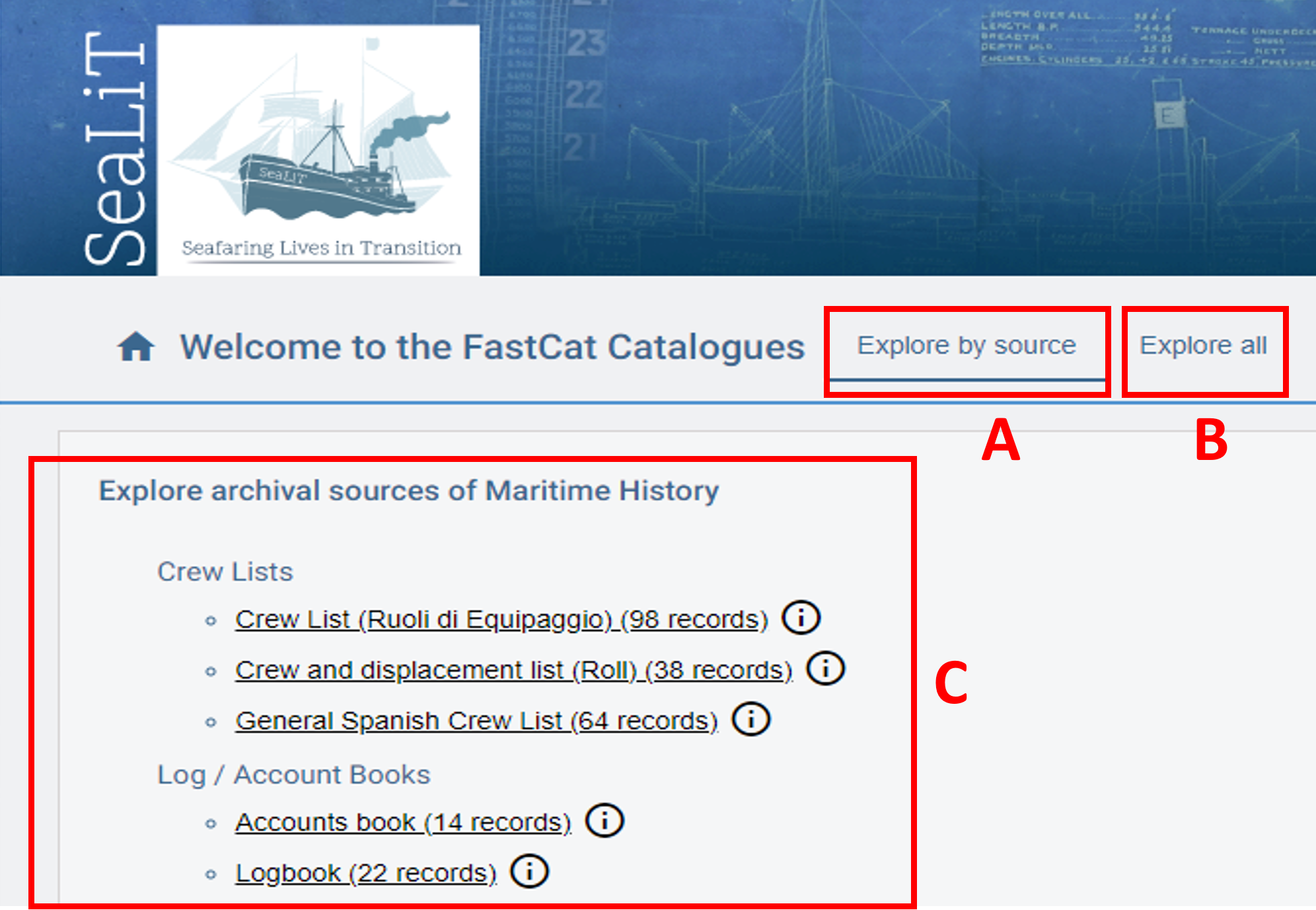}
    \caption{The home page of FastCat Catalogues as deployed for the case of SeaLiT.}
    \label{fig:screen1}
\end{figure}

Fig.~\ref{fig:screen1} shows the home page of FastCat Catalogues, as deployed for the case of the SeaLiT project. 
There are two tabs for exploring the archival data: 
\sq{Explore by Source} (Fig.~\ref{fig:screen1}-A)
and
\sq{Explore all} (Fig.~\ref{fig:screen1}-B). 
The former (default option) shows all types of sources (and their number of records) grouped in categories (Fig.~\ref{fig:screen1}-C), allowing the user to select one and start exploring its entities. 
The latter shows a set of categories of entities (like persons, locations, etc.), allowing users to explore entities across sources (Fig.~\ref{fig:screen3}-A). 
The data in both cases is dynamically loaded based on the configuration made to the application (more below).

\begin{figure*}[t]
    \centering
   \fbox{\includegraphics[width=17.3cm]{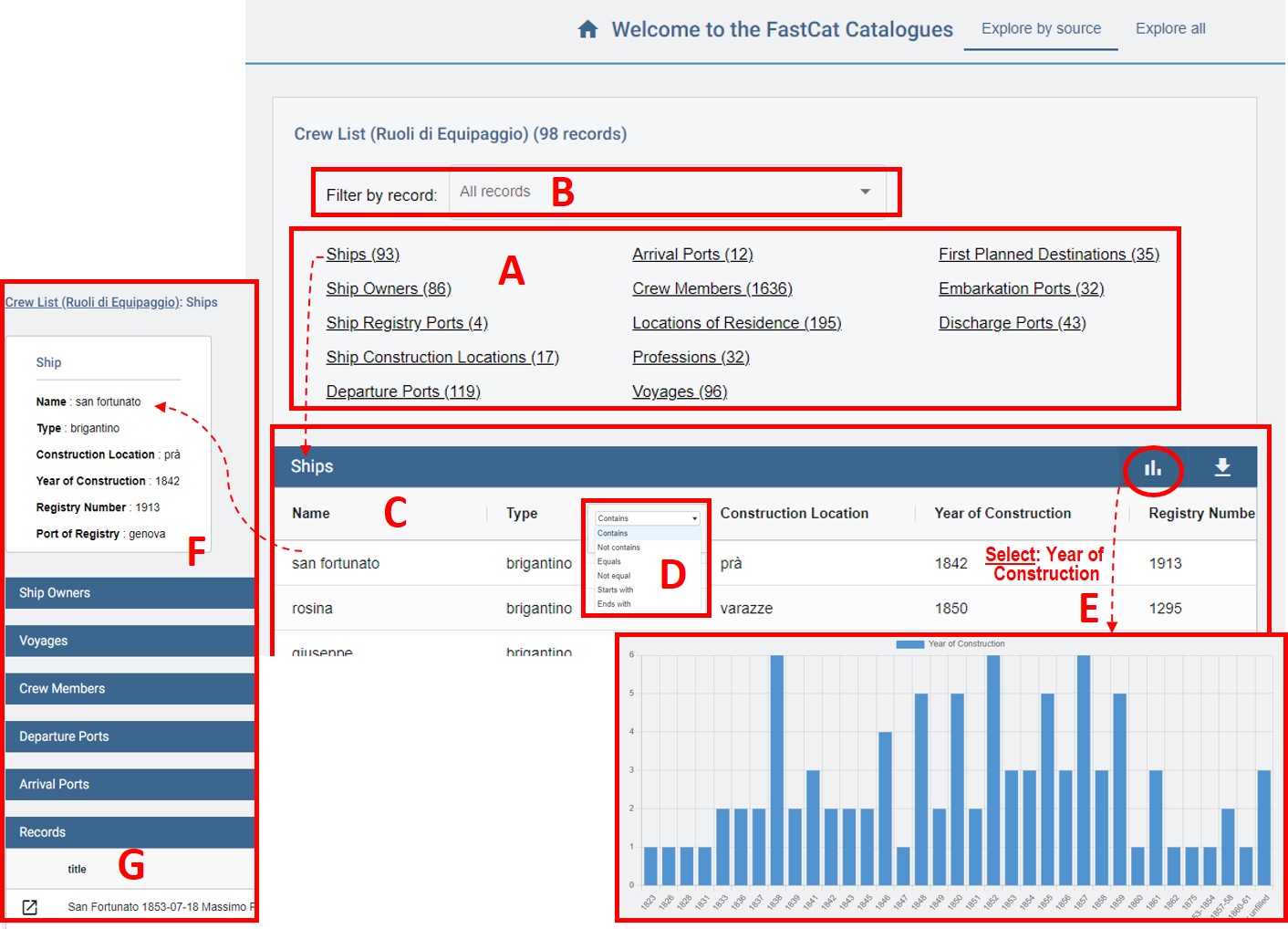}}
    \caption{Exploring the entities of a specific type of source and visualising aggregated information in a chart. }
    \label{fig:screen2}
\end{figure*}

\subsection{Explore by source} 
By selecting a specific type of source in the \sq{Explore by Source} tab, the user first gets an overview of the categories of entities (and their number) that exist in all records of this source type (Fig.~\ref{fig:screen2}-A). The user can also filter the displayed information by selecting a specific record (Fig.~\ref{fig:screen2}-B).
By selecting a category of entities, the user is shown with a table containing all instances of the selected category (Fig.~\ref{fig:screen2}-C). 
The user can filter the instances in the table by adding a filter in one or more of the table columns. Here there are different filtering options depending on the column type, e.g. \textit{contains, not contains, equals, not equal, starts with, ends with} for string values (Fig.~\ref{fig:screen2}-D). 
Also, the entity instances shown in a table can be grouped by selecting a specific column (corresponding to an entity property) and visualised in a chart (Fig.~\ref{fig:screen2}-E).

We should mention here that, if the original record does not contain a value for an entity property (e.g. there is no construction location for a ship), we display the value \textit{\sq{None or unfilled}} in the tables or charts. This is very important for researchers for getting valid information and making safe conclusions.

By selecting an entity instance in the table of entities, the user can inspect the connections of the selected entity with other entities and start browsing the related information (Fig.~\ref{fig:screen2}-F). In the case of SeaLiT, for instance, for a ship of the source type \sq{Crew List}, the user gets tables showing the ship's owners, voyages, crew members, departure ports, and arrival ports. Then, by selecting, for example, a departure port, the user can see all ships that have the same departure port together with their departure dates, and so on. In all these tables the user can apply filters,  group the entities by one of its properties and show a chart, or export the data in CSV format for further (offline) analysis.

For a selected entity, the user also gets a table with all records (transcripts) that mention the entity with the ability to directly visit them (Fig.~\ref{fig:screen2}-G).  

\subsection{Explore Across Sources}
\begin{figure*}
    \centering
   \includegraphics[width=15.5cm]{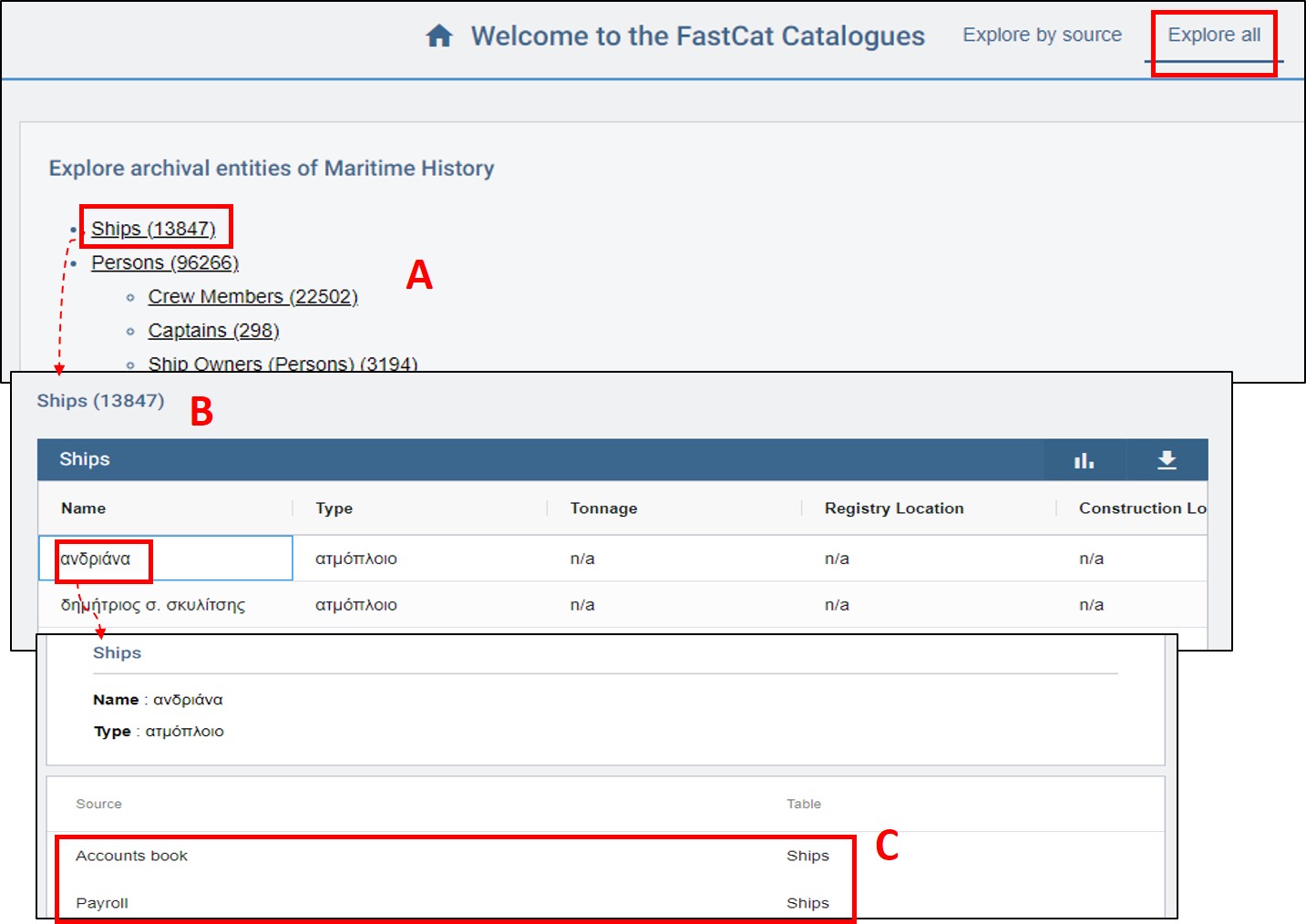}
    \caption{Exploring an entity across sources. }
    \label{fig:screen3}
\end{figure*}

The \sq{Explore all} tab allows users to explore the entities across all sources. The user is first shown a list with the different categories of entities mentioned in the transcripts of all sources (Fig.~\ref{fig:screen3}-A). 
By selecting a particular entity category (e.g. ships), the user is shown a table with all its instances (Fig.~\ref{fig:screen3}-B). The columns shown in the table for an entity category is the union of all the columns shown in the tables of the sources that contain the same entity category. 
If a source does not provide a value for a column, then we set \sq{n/a} in the corresponding cell. 

By selecting an entity instance from the table, the user is shown with all sources that mention the entity (Fig.~\ref{fig:screen3}-C). Then, by selecting one of the displayed sources,  the user is redirected to the corresponding \sq{Explore by source} page which shows the entity's connections  with other entities in the selected source.

\subsection{Technologies}
The front-end has been implemented in Angular\footnote{\url{https://angular.io/}} and the back-end in Express.js\footnote{\url{https://expressjs.com/}}, a Node.js framework for creating APIs.
For displaying the data in tables with filtering and ranking capabilities we use ag-grid\footnote{\url{https://ag-grid.com}}, while for visualising the table data in chart we use Chart.js\footnote{\url{https://chartjs.org}}.

\section{Configuration Model}
\label{sec:conf}

The application can be configured for use over any type (structure) of transcripts. It operates over a set of JSON files, each one containing the transcribed data of a single archival document. The JSON files are organised in folders, where each folder contains files of the same JSON structure (representing a specific archival source type).

For configuring the system, we first need to define the {\em templates.json} file. In this file, we provide for each different type of archival source: i) a category name (used for grouping the different types of sources), ii) an ID (used for creating the links to the original transcripts), iii) a name (shown in the UI), iv) a description (shown in the UI), and v) the name of its configuration file.

In the configuration file of each source, we define 
the entity categories (e.g. persons, ships, etc.) that appear in records of this source type and which will be available for exploration. For each entity category, we define the JSON fields that provide entity-related information like properties of the entity or its relations to other entities. 

Finally, for configuring the \sq{Explore all} functionality, we first need to define the names of all the supported entity categories and their grouping (in the \textit{explore\_all.json} configuration file). Each of the entity categories can then be configured by defining the sources and the tables in each source that provide instances (in the \textit{explore\_all\_conf.json} file). All other information needed for creating the entity tables is read from the source-specific configuration files.

This type of configuration allows the setup of  the application for use over any JSON structure, making it independent to the system used for data transcription. Details on how to prepare the configuration are available at the system's GitHub repository (Footnote~\ref{gitlink}).

\section{Evaluation and Use}
\label{sec:usecase}

\subsection{Requirements Satisfaction}
We evaluated the application in terms of how it can support finding answers to the four categories of real information needs described in Section~\ref{sec:reqAn}, as well as its ability to satisfy the requirement about the provenance of the displayed data. Specifically:

\textit{(i) Finding information about a particular entity.} 
The user can first select the category of the searching entity (Fig.~\ref{fig:screen2}-A or Fig.~\ref{fig:screen3}-A) and then detect the desired entity by using the table filtering options (Fig.~\ref{fig:screen2}-D). By selecting the desired entity, the user can explore information about it (properties, related entities, etc.) (Fig.~\ref{fig:screen2}-F).

\textit{(ii) Retrieving a list of entities based on one or more properties of these entities, together with additional information about them. }
The user can select a category of entity and then apply filters in the columns of the displayed table for defining the desired entity properties (Fig.~\ref{fig:screen2}-D).

\textit{(iii) Grouping a list of retrieved entities based on some property or characteristic. } 
As in (ii), the user can select a category of entity and then apply filters in the columns of the displayed table for defining the desired entity properties. Then, the entities can be grouped and visualised in a chart by selecting a specific entity property (Fig.~\ref{fig:screen2}-E).

\textit{(iv) Finding comparative information related to some entities.} 
The aggregated information displayed in a chart considers any applied filters. In this way, the user can inspect charts using different filters each time and thus find comparative information through such multiple interactions. For example, we can filter the list of a ship's crew members based on different residence locations and each time group the persons by their birth year (to check for any difference in the age of crew members for different residence locations).

\textit{(v) Finding the provenance of any piece of displayed information.} When the user selects an entity from the table of entities (Fig.~\ref{fig:screen2}-F), the last table called \q{Records} 
shows all records containing the displayed data of that particular entity (Fig.~\ref{fig:screen2}-G). The user can directly select a row and be redirected to the corresponding record.
Similarly, when the user explores entities across sources and selects an entity (Fig.~\ref{fig:screen3}-A,B), she/he is shown with all sources containing it (Fig.~\ref{fig:screen3}-C). By selecting one of the sources, the user is redirected in the corresponding \sq{Explore by source} page which shows the entity's connections with other entities and the corresponding records table.  

\subsection{Use in Maritime History Research} 
We have deployed the application in a real context (SeaLiT project) for supporting a large group of historians in exploring transcripts of their archival material. 
The archival documents studied by historians of SeaLiT consist of crew and displacement lists, logbooks, payrolls, account books, censuses, employment records, notarial deeds, and registers of different types such as sailors registers and naval ship registers. 
Archival documents covering these types of sources have been transcribed in tabular form by the historians (using the FastCat system~\cite{fafalios2021fast}) and stored in a JSON database.   
The total number of transcripts is more than 600, providing information for about 100K persons (sailors, etc.), 2.4K ships, 9.8K locations, and 1.1K legal entities (organisations).  
The application is publicly accessible for use.\footnote{\url{https://catalogues.sealitproject.eu/}}

\section{Conclusion and Future Work}
\label{sec:conclusion}

We have presented FastCat Catalogues, a web application that supports researchers and domain experts working with archival material in exploring and quantitatively analysing the transcribed data. 
The application is configurable, provenance-aware and goes beyond searching archival documents based on metadata or textual search on their full contents, by exploiting the interrelations of the entities mentioned in the documents. 

The system is currently used in maritime history research, for supporting historians in exploring the data of more than 600 records belonging to 20 different types of archival sources and which provide interrelated historical information for more than 100K entities (persons, ships, locations, etc.). 

A current issue is the fact that the same entity may appear under different representations in the archival documents. This can happen due to several reasons, like different language, unrecognisable characters, or difference in an entity's property. 
Our current work is concerned with the implementation of an effective and efficient solution to this \textit{entity 
 matching/resolution} problem~\cite{christophides2020overview}.
Another direction for future work is the \sq{semantification} of the application through the use of semantic technologies for the ontological representation of the data and their linking to external relevant resources.

\section*{Acknowledgements}

This work has received funding from the European Union’s Horizon 2020 research and innovation program under the Marie Sklodowska-Curie grant agreement No. 890861 (Project ReKnow), and the European Research Council (ERC) grant agreement No. 714437 (Project SeaLiT).

\bibliographystyle{ACM-Reference-Format}
\balance
\bibliography{MAIN_BIB} 

\end{document}